\documentclass[a4paper,showpacs,twocolumn,amsmath,amssymb,floatfix,prb,preprintnumbers,footinbib,unsortedaddress]{revtex4}
\usepackage[dvips]{graphicx}
\usepackage{dcolumn}
\usepackage{bm}
\usepackage{dsfont}
\usepackage{color}

\newcommand{\comment}[1]{}

\begin{document}


\title{Coherent Spin Ratchets} 

\author{Matthias Scheid}
\affiliation{Institut f{\"u}r Theoretische Physik, Universit{\"a}t Regensburg, D-93040, Germany} 
\author{Andreas Pfund}
\affiliation{Institut f{\"u}r Theoretische Physik, Universit{\"a}t Regensburg, D-93040, Germany} 
\affiliation{ETH H\"onggerberg, Solid State
Physics Lab, HPF E6, Z\"urich, CH-8093, Switzerland.}
\author{Dario Bercioux}
\email{dario.bercioux@physik.uni-freiburg.de}
\affiliation{Institut f{\"u}r Theoretische Physik, Universit{\"a}t Regensburg, D-93040, Germany} 
\affiliation{Physikalisches Institut, Albert-Ludwigs-Universit\" at, D-79104 	Freiburg, Germany}
\author{Klaus Richter} 
\affiliation{Institut f{\"u}r Theoretische Physik, Universit{\"a}t
Regensburg, D-93040, Germany} 

\date{\today}

\begin{abstract}
We demonstrate that the combined effect of a spatially periodic
potential, lateral confinement and spin-orbit interaction gives rise
to a quantum ratchet mechanism for spin-polarized currents in
two-dimensional coherent conductors. Upon adiabatic ac-driving,  in
the absence of a net static bias, the system generates a directed spin
current while the total charge current is zero. We analyze the
underlying mechanism by employing symmetry properties of the
scattering matrix and numerically verify the effect for different
setups of ballistic conductors. The spin current direction can be
changed upon tuning the Fermi energy or the strength of the 
Rashba spin-orbit coupling.
\end{abstract} 

\maketitle

\section{Introduction}
Charge transport is usually studied by considering current in response
to an externally applied bias. However, there has been growing
interest throughout the last decade in mechanisms enabling directed
particle motion in nanosystems without applying a net dc-bias. In this
respect, ratchets, periodic structures with broken spatial symmetry,
\emph{e.g.}\ saw tooth-type potentials, represent a prominent class.
Ratchets in the original sense are devices operating far from
equilibrium by converting thermal fluctuations into directed particle
transport in the presence of unbiased time-periodic
driving~\cite{Reimann02}. First discovered in the context of
(overdamped) classical Brownian motion~\cite{Haenggi96,Juelicher97},
the concept of dissipative ratchets was later generalized to the
quantum realm~\cite{Reimann97}.  More recently, coherent ratchets and
rectifiers have gained increasing attention. They are characterized by
coherent quantum dynamics in the central periodic system in between
leads where dissipation takes place. Proposals
com\-prise molecular wires~\cite{Lehmann02} and cold atoms in optical
lat\-ti\-ces \cite{Lundh05}, besides Hamiltonian ratchets~\cite{Flach00}.
Experimentally, ratchet-induced charge flow in the coherent regime was
first observed in a chain of triangular-shaped lateral
quantum dots \cite{Linke99}
and later in lateral superlattices~\cite{Hoehberger01}.

Here we propose a different class of ratchet devices, namely 
{\em spin ratchets} which act as sources for spin currents with
simultaneously vanishing charge, respectively particle currents. To be
definite we consider coherent transport through ballistic mesoscopic
conductors in the presence of spin-orbit (SO) interaction~\cite{zeeman}. Contrary to
particle ratchets, which rely on
asymmetries in either the spatially periodic modulation or the
time-periodic driving, a SO-based ratchet works even for symmetric
periodic potentials.  As possible realizations we have in mind 
semiconductor heterostructures with 
Rashba SO interaction~\cite{rashba:1960} that can be tuned in
strength by an external gate voltage allowing 
to control the spin evolution.

Among other features it is this property which is triggering recent
broad interest in semiconductor-based spin
electronics~\cite{zutic:2004}.  Also since direct spin injection from
a ferromagnet into a semiconductor remains
problematic~\cite{schmidt:2000}, alternatively, several suggestions
have been made for generating spin-polarized charge carriers without
using magnets. In this respect, spin pumping appears pro\-mising,
\emph{i.e.}\ the generation of spin-polarized currents at zero bias via
cyclic variation of at least two parameters. Different theoretical
proposals based on SO~\cite{Sharma03} and Zeeman~\cite{Mucciolo02}
mediated spin pumping in non-magnetic semiconductors have been put
forward~\cite{Tserkovnyak05} and, in the latter case, experimentally
observed in mesoscopic cavities~\cite{Watson03}.

While pumps and ratchets share the appealing property of generating
directed flow without net bias, ratchet transport requires only a
single driving parameter, the periodic ratchet potential has a strong
collective effect on the spin current and gives rise to distinct
features such as spin current reversals upon parameter changes.

\section{Model and symmetry considerations}
We consider a two-dimensional coherent ballistic conductor in the 
plane ($x,z$) connected to two nonmagnetic leads. The
Hamiltonian of the central system in presence of Rashba SO interaction 
reads
%
%
\begin{equation}\label{hamiltonian}
\mathcal{H}_\text{c} = \frac{\hat{p}^2}{2m^*} + 
\frac{\hbar k_\text{SO}}{m^*} (\hat{\sigma}_x \hat{p}_z -
\hat{\sigma}_z \hat{p}_x)  + U(x,z) \, .
\end{equation}
%
%
Here $m^*$ is the effective electron mass, $U(x,z)$ includes the
ratchet potential in $x$- and a lateral transverse confinement in
$z$-direction, and $\hat{\sigma}_i$ denote Pauli spin matrices.  The
effect of the SO coupling with strength $k_\text{SO}$ is twofold: it
is leading to spin precession 
and it is coupling transversal modes in the confining
potential~\cite{mireles:2001}.

In view of a ratchet setup we consider an additional time-periodic
driving term $\mathcal{H}_V(t)$ due to an external bias potential
$V(t)$ with zero net bias (rocking ratchet). 
We study adiabatic driving (such that the system can adjust to the 
instantaneous equilibrium state), assuming that the driving period $t_0$ 
is large compared to the relevant time scales for 
transmission. This is the case in
related experiments~\cite{Linke99}.  
The entire Hamiltonian then reads 
%
%
\begin{equation}\label{driving}
\mathcal{H}= \mathcal{H}_\text{c} + \mathcal{H}_V(t) \quad ; \quad
\mathcal{H}_V(t)=V(t)g(x,z;V) \, ,
\end{equation}
%
%
where $g(x,z;V)$ describes the spatial distribution 
of the voltage drop and should in principle be obtained
self-consistently from the particle density.

We model spin-dependent transport within a scattering approach
assuming that inelastic processes take place only in the reservoirs.
Then the probability 
amplitude for an electron to pass through the conductor is given 
by the scattering matrix
$\mathcal{S}_{n\sigma;n'\sigma'}(E,V)$, where 
$n',n$ denote transverse modes and $\sigma' ,\sigma =\pm1$ 
the spin directions in the incoming and outgoing lead, respectively. 
Making use of the unitarity of the scattering matrix,
$\mathcal{S}\mathcal{S}^\dag= \mathcal{S}^\dag\mathcal{S} = \mathds{1}$,
and summing over all open channels in the left (L) and right (R) lead, respectively, 
we find the relations
%
%
\begin{equation}\label{Sunit}
\!\!
\sum_{\substack{n,\sigma \in \text{R}\\n',\sigma ' \in \text{R}\cup\text{L}}}
\!\!\!\!\!\left| \mathcal{S}_{n,\sigma ;n',\sigma '}\right| ^2=\!\!\!
\sum_{n,\sigma \in \text{R}}\! 1 \; , \!\!\!\!\!\!
\sum_{\substack{n,\sigma \in \text{R}\\n',\sigma ' \in \text{R}\cup\text{L}}}
\!\!\!\!\!\!\!\sigma\left| \mathcal{S}_{n,\sigma ;n',\sigma '}\right| ^2=0 \, .
\end{equation}
%
%

For the further analysis, we consider an unbiased square wave driving 
$V(t)=V_0 \,\text{sign} \left[ \sin (2\pi t/t_0) \right]$, restricted to the values $\pm V_0$ ($V_0>0$);
generalizations to, \emph{e.g.}, harmonic driving are straight forward.  The
ratchet current is then given by the average of the steady-state currents
in the two opposite rocking situations, $\langle I(V_0) \rangle =~
[I(+V_0)+I(-V_0)]/2$, which we compute within the Landauer
formalism relating conductance to transmission.

Contrary to charge current, spin current is usually not conserved.
Thus it is crucial to fix the measuring point, which we choose to be 
inside the right lead. 
Then, in view of Eq.~(\ref{Sunit}), the ratchet charge $\langle
I_\text{C} \rangle$ and spin $\langle I_\text{S} \rangle$ currents can
be expressed as
%
%
\begin{equation}\label{current}
 \langle I_{\text{C/S}}(V_0) \rangle = G_{\text{C/S}} \!
\int_{E_\text{C}}^{\infty} \!\! \text{d}E \, 
\Delta f(E,V_0)\Delta T_{\text{C/S}}(E,V_0) \, .
\end{equation}
%
%
Here, the prefactor $G_{\text{C/S}}$ is equal to $e/2h$ for the 
charge current and $1/8\pi$ for the spin current. 
$E_C$ denotes the energy of the 
conduction band edge, $\Delta f(E,V_0)= \left[ f(E,E_\text{F}+V_0/2) -
f(E,E_\text{F}-V_0/2) \right]$ is the difference between the 
Fermi functions in the leads, and 
%
%
\begin{equation}\label{ratchet:transmission}
\Delta T_{\text{C/S}}(E, V_0)=T_{\text{C/S}}(E,+V_0) -
T_{\text{C/S}}(E,-V_0).
\end{equation} 
%
%
With $ T_{\sigma,\sigma'}  \! = \! \sum_{ n \in \text{R},n' \in \text{L}} \left| \mathcal{S}_{n,\sigma;n',\sigma'}\right|^2$,
the transmission probabilities for charge and spin in (\ref{ratchet:transmission}) are defined as
%
%
\begin{eqnarray}
\label{transmission-C}
T_\text{C}(E,V)\! &=& 
\sum_{\substack{\sigma'=\pm 1 \in \text{L} \\ \sigma = \pm 1 \in \text{R} }}
T_{\sigma,\sigma'}(E,V) \, , \\
\label{transmission-S}
T_\text{S}(E,V)\! &=& \!\!\! 
\sum_{\sigma'=\pm 1 \in \text{L}}\!\! \left[ T_{+,\sigma'}(E,V)
  - T_{-,\sigma'}(E,V)\right] .
\end{eqnarray}
%
%
The latter is given by the difference between the transmission of
spin-up and spin-down electrons upon exit, with the spin measured with
respect to the $z$-axis.

%
%
\begin{figure}[t]
 \centering \includegraphics[width=0.9\columnwidth]{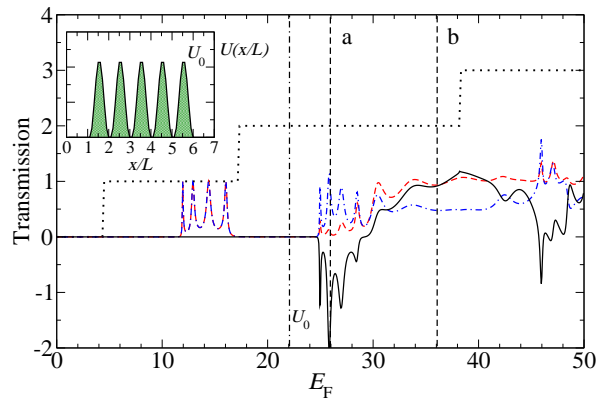}
	\caption{(\emph{Color online}) Spin-dependent transmissions as
	a function of the injection energy $E = (kL)^2$
	in the presence of Rashba spin-orbit interaction 
	($k_\text{SO}L=1.5$) for a short
	periodic chain of five symmetrical potential barriers (see inset, barrier
	height $U_0=22$) and moderate rocking amplitude $V_0=2$.  
	The dashed (red) and dotted (blue) lines indicate $T_\text{S}$, 
	Eq.~(\ref{transmission-S}), in the two
	rocking situations. The solid (black) line depicts
	the ratchet spin transmission, Eq.~(\ref{ratchet:transmission}),
	the sign indicating the flow direction. 
	For reference, the dashed-dotted (green) curve shows 
	 $T_\text{C}$, Eq.~(\ref{transmission-C}),
	 and the staircase 
	function  $T_\text{C}$ for a wire without
	potential barriers and SO interaction.  \label{figure:one}}
\end{figure}
%
%
%
%
\begin{figure*}
\centering \includegraphics[width=0.7\textwidth]{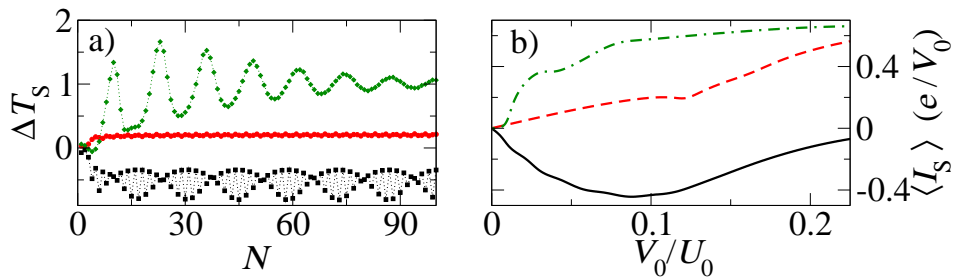}	
	\caption{(\emph{Color online}) (a) Ratchet spin
	transmission as a function of the number of barriers $N$ for
	$k_\text{SO}L=1.5$, $U_0=22$, $V_0 = 2$, 
	and energies $E=(kL)^2=$24 (black symbols, lower line), 33 (red, middle line) and 
	35.5 (green, upper green).
	 (b) Ratchet spin conductance $\langle I_{\rm S} \rangle  (e/V_0)$ 
	at zero temperature in units of $e G_\text{S}$ 
	as a function of applied voltage $V_0$ for 
	$N=20$, $k_\text{SO}L=1.5$, $U_0=22$ and $E=$24 (black solid line), 33 
	(red dashed line) and 35.5 (green dash-dotted line).
	 \label{figure:two}}
\end{figure*}
%
%

Equation (\ref{ratchet:transmission}) indicates that 
$\Delta T_{\text{C/S}}(E, V_0)$, and thereby the average conductance,
vanishes in the linear response limit $V_0 \rightarrow 0$. In the
following we consider the nonlinear regime and devise a minimum model for 
a spin ratchet mechanism by assuming identical leads  and 
a spatially symmetric potential $U(x,z)$ in Eq.~(\ref{hamiltonian}). 
The total Hamiltonian~(\ref{driving}) is then invariant under 
the symmetry operation 
$\hat{\mathcal{P}}=\hat{\mathcal{C}}\hat{R}_x\hat{R}_V\hat{\sigma} _z$,
where $\hat{\mathcal{C}}$ is the operator of complex 
conjugation, $\hat{R}_x$ inverses the $x$-coordinate and $\hat{R}_V$ changes the 
sign of the applied voltage $(\pm V \leftrightarrow \mp V)$.
The action of $\hat{\mathcal{P}}$ on the scattering states is to
switch between the two rocking situations and to 
exchange the leads, \emph{i.e.}, a mode index $n$ is replaced by its
corresponding mode $\tilde{n}$. Moreover, incoming
(out\-going) states are transformed into outgoing (incoming) states
with complex conjugated amplitude.
It is then straightforward to show that 
%
%
\begin{equation}\label{s:simmetry}
S_{n,\sigma;n',\sigma '}(E,\mp V_0)=\sigma\sigma 'S_{\tilde{n}',\sigma ';\tilde{n},\sigma}(E,\pm V_0)\; ,
\end{equation}
%
%
leading to a vanishing charge current $\langle I_\text{C} (V_0) \rangle$ and 
a simplified expression for the ratchet spin transmission
(\ref{ratchet:transmission}):
%
%
\begin{equation}\label{new:spin:current}
\Delta T_{\text{S}}(E,V_0)=2 \left[
T_{+,-}(E,+V_0)-T_{-,+}(E,+V_0)\right] \, .
\end{equation}
%
%

\section{Ratchet mechanism: numerical results}
We illustrate the prediction for a ratchet spin current (Eq.~(\ref{current})
with (\ref{new:spin:current})) by performing numerical
calculations for the Hamiltonian
(\ref{hamiltonian},\ref{driving}). The amplitudes
$\mathcal{S}_{n\sigma^\prime;m\sigma}(E,V)$ are obtained by projecting
the Green function of the open ratchet system onto an appropriate set
of asymptotic spinors defining incoming and outgoing channels.  For
the efficient calculation of the $\mathcal{S}$-matrix elements a
real-space discretization of the Schr\"odinger
equation combined with a recursive algorithm for the
Green functions was implemented for spin-dependent
transport~\cite{frustaglia,bercioux07}.

As a model for a spin ratchet we consider a ballistic two-dimensional
quantum wire of width $W$ with Rashba SO strength $k_\text{SO}$ and a
one-dimensional periodic modulation (period $L$) composed of a set of
$N$ symmetric potential barriers $U(x) \! =\! U_0 [1-\cos(2\pi x/L)]$. We
assume a linear voltage drop \cite{voltage-drop} across the system,
$g(x,z) = 1/2 -x/(NL)$ in Eq.~(\ref{driving}).
To simplify the assessment of the rich parameter space 
 ($E_F, U(x), V, k_\text{SO}, N$) of the problem ($L$ can be
scaled out and $W$ is fixed to $1.5L$) and to analyze the mechanisms for spin currents, we first
consider a strip with $N\!=\!5$ potential barriers (see
inset in Fig.~\ref{figure:one}) and few open transverse modes.
Figure~\ref{figure:one} shows the
numerically obtained spin transmission probabilities $T_\text{S}(E,V)$,
Eq.~(\ref{transmission-S}), for $k_\text{SO}L = 1.5$
in the two rocking situations $\pm V_0$ (dashed and dotted line,
respectively). The solid line represents
the resulting ratchet spin transmission $\Delta T_{\rm S}$,
Eq.~(\ref{ratchet:transmission}).
For comparison, the dashed-dotted curve shows 
$T_{\rm C}(+V_0) \!=\! T_{\rm C}(-V_0) $, 
Eq.~(\ref{transmission-C}), and the staircase function
the successive opening of transverse modes $n = 1,2,3$ in the overall
transmission of the conductor without potential barriers and 
SO interaction.

%
%
\begin{figure}[b]
 \centering \includegraphics[width=\linewidth]{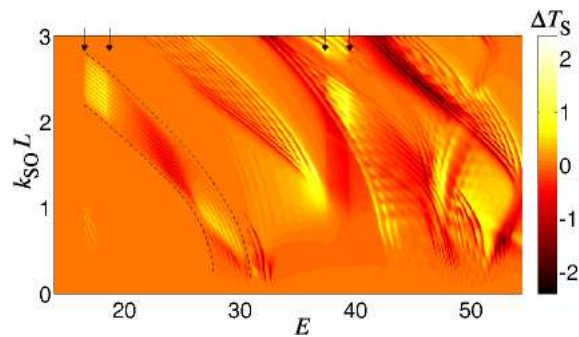}
	\caption{(\emph{Color online}) Ratchet spin transmission as a function 
	of energy $E\!=\!(kL)^2$ and 
	SO interaction $k_\text{SO}L$ for $N=20$, $V_0=2$ and $U_0=22$. The dashed lines are 
	a guide to the eye for the shift of the first Bloch band.
	 \label{figure:two:bis}}
\end{figure}
%
%
At energies below $U_0$ and within the first conducting transverse mode the
spin transmissions $T_\text{S}(\pm V_0)$ are zero, while the total transmission
$T_\text{C}(\pm V_0)$ is suppressed up to a sequence of four peaks
representing resonant tunneling through states which can be viewed as
precursors of the lowest Bloch band in the limit of an infinite
periodic potential. When the second mode is opened spin polarization
is possible (see model below) and takes different values in the two rocking
situations leading to a finite ratchet spin transmission. Two
transmission peak sequences, related to the lowest one, 
reappear at higher energies (around $E\!=\!$ 24 and 45), both for
$T_\text{C}(\pm V_0)$ and for $T_\text{S}(\pm V_0)$, owing to
corresponding resonant Bloch states involving the second and third
transverse mode. The enhanced ratchet spin transmission at the opening of the 
third mode (at $E=38$) can be associated to a 'classical' rectification
effect resulting from a different number of open modes in one lead 
in the two rocking situations. 

Figure~\ref{figure:one} demonstrates moreover that the
associated spin current changes sign several times upon variation of
the energy, opening up the experimental possibility to control the
spin current direction through the carrier density via an external
gate.  This energy dependence of the spin current implies also current
inversion as a function of temperature~\cite{bercioux07}.  Such
behavior is considered as typical for quantum (particle)
ratchets~\cite{Reimann97,Linke99}.

In Fig.~\ref{figure:two}(a) we present the ratchet spin
transmission  $\Delta T_\text{S}$ as a function of the barrier number $N$.
Obviously,  $\Delta T_\text{S}$ approaches different asymptotic values depending 
on the Fermi energy: For energies in resonance with the first Bloch band
(lowest trace), $\Delta T_\text{S}$  exhibits a 
long- and a short-scale frequency 
oscillation owing to commensurability between the spin precession length
$L_\text{SO} \!=\! \pi / k_\text{SO}$ and the geometry of the periodic
system. For off-resonant injection energies two characteristic, distinct 
behaviors are shown: a large-scale oscillation (upper curve) and a nearly constant
behavior (middle trace), respectively. It is remarkable that 
in all cases the periodic structure enhances
considerably the absolute value of $\Delta T_\text{S}$.

In Fig.~\ref{figure:two}(b) we show the ratchet spin conductance,
$\langle I_{\rm S} \rangle (e/V_0)$, as
a function of the applied driving voltage for a system with 20
barriers. For energies within the first Bloch band (solid line), the
ratchet spin conductance exhibits a non-monotonic behavior.
For the off-resonant cases (dashed and dashed-dotted line)
it is monotonically increasing in the voltage window considered.

In Fig.~\ref{figure:two:bis} we present the ratchet spin
transmission as a function of injection energy $E$ and 
Rashba SO interaction $k_\text{SO}$. We find a rich 
structure in the explored parameter space, where both large positive 
and negative values of the ratchet spin transmission can be observed. 
In the whole energy range peaks due to resonant tunneling are 
visible, which are shifted to lower energies for increasing 
SO coupling (\emph{e.g.}, region between dashed lines). 
Furthermore, we observe discontinuities in the spin transmission 
at energies where an additional transversal mode 
in one of the leads opens up (marked by arrows).

For InAs quantum wells $L_\text{SO}$ is of the order of 0.2~$\mu$m~\cite{expSO}, in 
InGaAs it has been tuned from 0.7 
to 1.6~$\mu$m~\cite{expSO2} and in GaAs from 2.3 to 5.6~$\mu$m~\cite{zumbuhl:2002}; 
the range of SO coupling $k_\text{SO} L \!=\! \pi L / L_\text{SO}$ given in 
Fig.~\ref{figure:two:bis} can be achieved in experiments for period $L$ on scales of 
$\mu$m. Spin-polarized currents as predicted
here exceed those observed with experimental detection schemes, reported,
\emph{e.g.}, in Ref.~\onlinecite{Watson03}.

\section{Ratchet mechanism: simplified model}
\begin{figure}
	\centering \includegraphics[width=0.8\columnwidth]{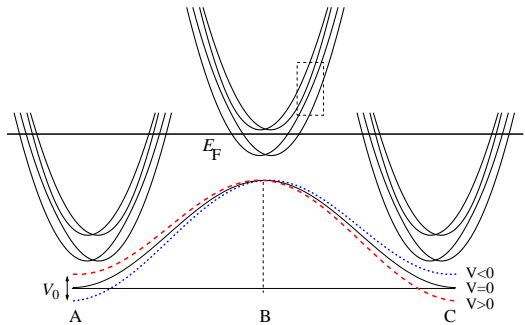}
	\caption{(\emph{Color online}) Illustration of the spin
	polarization mechanism for transmission through a strip with a single 
	adiabatic symmetric potential barrier $U(x)$ (solid line) in the two rocking situations
	(dashed and dotted line). At points A,B and C the 
	position-dependent energy dispersion relation $E_n(k_x)$ is sketched
	with respect to the Fermi energy $E_\text{F}$ (horizontal line) for two 
	transverse modes and SO-induced spin splitting of each mode. 
	\label{fig:barrier}}
\end{figure}

Finally we present a simplified model providing additional insight into the underlying 
mechanism for the occurrence of a finite ratchet spin current.
We consider a wire with two open transverse modes $(n=1,2)$ and a smooth symmetric 
potential barrier $U(x)$ in the two rocking situations, see Fig.~\ref{fig:barrier}.
%
%
%
%
Upon adiabatically traversing the barrier from A via B to C, the spin-orbit split 
energy spectrum $E_n(k_x)$ for electrons is shifted up and down.
For fixed Fermi energy  $E_\text{F}$, the initial shift causes a 
depopulation of the upper levels (n=2) and a spin-dependent repopulation while moving from B to C. 
When $E_\text{F}$ is traversing an anti-crossing between successive 
modes (see the region indicated by the dashed window in 
Fig.~\ref{fig:barrier}), there is a certain probability $P$ for the
electrons to change their spin state. This causes an asymmetry between
spin-up and -down states for the repopulated levels~\cite{eto:2005}.
The related transition probability can be computed in a Landau-Zener picture
and reads, for a transverse parabolic confinement of frequency $\omega_0$, 
%
%
\begin{equation}
P(\pm V_0)\!=\! 1-\exp \left\{\frac{-\pi k_\text{SO} \omega_0/  \Sigma_z}{
(\partial/\partial x) [U(x,z)\pm V_0 g(x,z)]} \right\} \, .
\label{landau-zener}
\end{equation}
%
%
Here $\Sigma_z$ denotes the difference in the polarizations of the two modes 
involved. 
The spin transmission is proportional to $P(V)$ and thus different 
in the two rocking situations. Hence, the ratchet spin 
current $\langle I_\text{S} (V_0)\rangle$ is nonzero, even in the case of
a symmetric barrier. Expanding Eq.~(\ref{landau-zener}) for
small $V_0$ allows to qualitatively understand the linear dependence of 
the ratchet spin conductance for small $V_0$ in Fig.~\ref{figure:two}. 
However, a quantitative explanation of the spin ratchet effect 
for a periodic, non-necessarily adiabatic potential is beyond this
model.

\section{Conclusions}
The overall analysis indicates that the ratchet setup, carrying
features of a spin rectifier, differs from the
proposals~\cite{Sharma03,Mucciolo02,Watson03} for spin pumps, since it
operates with a single driving parameter, invokes quantum tunneling
effects, and the spin transmission is governed by the spatial 
periodicity of the underlying potential.
Further calculations~\cite{bercioux07} for combined Rashba- and 
Dresselhaus~\cite{Dresselhaus} SO coupling do not alter the overall 
picture but show that the spin current direction can be changed upon 
tuning the relative strength of the two coupling mechanisms.

To summarize, we showed that ratchets built from mesoscopic conductors
with SO interaction generate spin currents in an
experimentally accessible parameter regime. Many further interesting
questions open up within this new concept, including the exploration
of spin ratchet effects for non-adiabatic driving and for dissipative
and non-equilibrium particle and spin dynamics.

{\em Acknowledgements:} 
We thank P. H\" anggi, M.~Grifoni and
M. Strehl for useful discussions and acknowledge support from the
German Science Foundation (DFG) within SFB 689.


\begin{thebibliography}{99}
\bibitem{voltage-drop} 
An analysis \cite{bercioux07} of different models for the voltage drop
shows that the results for the spin current, up to slight quantitative
changes, are not altered qualitatively.
\end{thebibliography}
\end{document}